 \documentclass[final,3p,times,twocolumn]{elsarticle}

 \usepackage{graphicx}
 \usepackage{epsfig}

\usepackage{amssymb}
\usepackage{amsmath}

\begin{document}

\begin{frontmatter}



\title{Least-action principle and path-integral for classical mechanics}


\author[2]{E. Cattaruzza}
\author[1,2]{E. Gozzi}
\author[3]{A. Francisco Neto }

\address[1]{ Department of Physics (Miramare campus),
University of Trieste, Strada Costiera 11, 34014 Trieste, Italy.}
\address[2]{Istituto Nazionale di Fisica Nucleare, Sezione di Trieste, Italy.}
\address[3]{DEPRO. Escola de Minas, Campus Morro do Cruzeiro, UFOP 35400-000 Ouro-Preto, MG, Brazil.}

\begin{abstract}
In this paper we show how the equations of motion of a superfield, which makes its appearance  in a path-integral 
approach to classical mechanics, can be derived without the need of the least-action principle.
\end{abstract}


\end{frontmatter}
\section{Introduction}
In the 30's Koopman and von Neumann  (KvN) gave an operatorial approach to classical mechanics (C.M.) \cite{Koopman}. In the late 80's a functional (path-integral) counterpart of the KvN was proposed \cite{gozzireuter}.
 To distinguish this path-integral from the quantum one we will use the acronym CPI (classical path-integral) for the first and QPI for the second.
 The difference is that while the QPI gives a weight $\exp\left( \frac{i}{\hbar}\,S[q(t)]\right)$ to every path, the CPI gives weight one to the classical paths and weight zero to the others.\par  Indicating with $\varphi^a$ the points  in phase space
 \[
 \varphi^a=(q^1,\dots,q^n;p^1,\dots,p^n)
 \]
 and with $H(\varphi^a)$ the Hamiltonian, the eqs. of motion are:
\begin{equation}
 {\dot\varphi}^a=\omega^{ab}\frac{\partial H}{\partial \varphi^b},\label{E.C.M.}
 \end{equation}
 where $\omega^{ab}$ is the symplectic matrix \cite{mardsen}. The CPI has a generating functional of the form:
\begin{equation}
\mathcal Z[0]=\int \mathcal D \varphi^a\,\tilde{\delta}\left[\varphi^a-\phi_{cl}^a(\phi_0,t_0;t)\right]
\label{CPI}\end{equation}
 where $\phi_{cl}^a(\phi_0,t_0;t)$ are the solutions of the eq. of motion (\ref{E.C.M.}) with some initial condition $\phi_0$. The integration $\mathcal D \varphi^a$
is the functional integration, including an integration overall possible initial configurations $\varphi_0^a$. The $\tilde{\delta}[\dots]$ 
 is a functional Dirac-delta which gives weight "one" to the classical solutions and "zero" to all the others.
 We can rewrite (\ref{CPI}) as 
\begin{align}
\mathcal Z[0]&=\int \mathcal D \varphi^a\,\tilde{\delta}\left[\varphi^a-\phi_{cl}^a(\phi_0,t_0;t)\right]=\label{CPIGenerator}\\
& =\int \mathcal D \varphi^a\,\delta\left[\dot{\varphi}^a-\omega^{ab}\frac{\partial H}{\partial \varphi^b}\right]\,\textrm{det}\left(\delta^a_b\,\partial_t-\omega^{al}\frac{\partial^2H}{\partial\varphi^l\varphi^b}\right).\nonumber
\end{align}
 Let us perform a Fourier transform of the Dirac delta appearing in (\ref{CPIGenerator}) introducing $2n$ extra variables $\lambda_a$ and let us exponentiate the determinant using a set of $4n$ Grassmann variables $c^a$ and $\bar c_a$ (with $a=1,\dots,2n$). The final result is the following:  
\begin{equation}
\mathcal Z[0]=\int \mathcal D \varphi^a\mathcal D \lambda_a\mathcal D c^a\mathcal D \bar{c}_a\,\textrm{exp}\left[{i\int dt \tilde{ \mathcal L}}\right]\label{FullCPIGeneratingFunctional}
\end{equation}
where 
\begin{equation}
\tilde{\mathcal L} = \lambda_a\,\dot\varphi^a+i\bar c_a\,\dot c^a-\lambda_a\,\omega^{ab}\partial_bH-i\bar c_a\omega^{ad}\partial_d\partial_bH\,c^b.
\label{CPILagrangian}
\end{equation}
In this formulation of C.M. we passed from $2n$ variables $\varphi^a$ to $8n$ ones which are $(\varphi^a,\lambda_a,c^a,\bar c_a)$ 
and from the standard lagrangian of C.M.
\begin{equation}
 L=\frac{1}{2}\dot \varphi^a\omega_{ab}\varphi^b-H(\varphi^a) \label{C.M.Lagrangian}
\end{equation}
to the complicated $\tilde{\mathcal L}$  of eq.(\ref{CPILagrangian}).  Actually it is easy to simplify the formulation contained in (\ref{FullCPIGeneratingFunctional}) and (\ref{CPILagrangian}) and it was done in \cite{gozzireuter} and \cite{geomdeq}. The trick is to introduce two grassmaniann partners of time, which we will indicate with $\theta, \bar \theta$. Next let us define the following functions of $t,\theta,\bar\theta$:
\begin{equation} 
 \Phi^a(t,\theta,\bar{\theta})\equiv \varphi^a(t)+\theta\,c^a(t)+\bar{\theta}\,\omega^{ab}\,\bar{c}_b(t)+i\,\bar{\theta}\theta\,\omega^{ab}\lambda_b(t).\label{SuperField}
\end{equation}
This is a sort of multiplet which put together all the $8n$ variables ($\varphi^a,\lambda_a,c^a,\bar c_a$) and it is known in supersymmetric jargon as "superfield".
It is easy to prove that:  
\[
\tilde{\mathcal L}(\varphi,\lambda,c,\bar{c})=i\int d\theta\,d\bar{\theta}\,L\left[\Phi\right]+\textrm{(s.t.)}\,,\label{CPIQPILagrangian}
\]
where $\tilde{\mathcal{L}}$ is the lagrangian in (\ref{CPILagrangian}) while $L$ is the lagrangian in (\ref{C.M.Lagrangian}) but with the superfield $\Phi^a$ replacing $\varphi^a$
and (s.t.) is a surface term. The generating functional (\ref{FullCPIGeneratingFunctional}) can then be written as 
\begin{equation}
\mathcal Z[0]=\int \mathcal D\Phi\,\exp\left[i\int i\,d\theta\,d\bar{\theta}\,L[\Phi]\right], \label{SuperCPI}
\end{equation}
which strongly resembles the one of the QPI. For more details see ref.\cite{geomdeq}.
\section{Equation of motion}
In ref.\cite{gozzireuter} and \cite{geomdeq} we wrote  that the equations of motion for the $8n$ variables could be derived from $\tilde{\mathcal L}$ of eq.(\ref{CPILagrangian}) or, via the superfield, from $L[\Phi^a]$. 
We somehow assumed  a least-action principle for the lagrangians $\tilde{\mathcal L}(\varphi,\lambda,c,\bar c)$ or $L\left[\Phi\right]$. In this paper we will show that there is actually no reason to assume this principle which usually  holds only for $L[\varphi^a]$ and provides the correct equation for only $\varphi^a$. Using this principle for $\tilde{\mathcal L}(\varphi,\lambda,c,\bar c)$  we got the eqs. of motion not only for $\varphi^a$ but also for $\lambda_a,\,c^a,\,\bar c_a$. They are: 
\begin{align}
&\dot\varphi^a-\omega^{ab}\partial_bH=0\label{PhaseSpaceEqn}\\
&\dot c^a-\omega^{ab}\partial_{b}\partial_{k}H\,c^k=0\nonumber\\
&\dot{\bar{c}}_a+\omega^{ed}\partial_{a}\partial_{d}H{\bar c}_e=0\nonumber\\
&\dot \lambda_b+\omega^{ed}\partial_{b}\partial_{d}H\lambda_e+i\,\omega^{fe}\partial_{b}\partial_{d}\partial_{e}H\,\bar c_f\,c^d=0\nonumber.
\end{align}
While the eqs. we got for $\varphi^a$ are the correct ones, no one can guarantee that this is the case for $c^a,\bar c_a,\,\lambda_a$. These equations should be derived using only the path-integral (CPI) and then compared with those obtained from the least action principle. After all we have only the path-integral as a tool and we should only use that. The reader may think, like it is done in the QPI, that in the path-integral one could use a saddle point technique to get the eqs. of motion. This can actually   
be done in the QPI where we have a small-parameter, $\hbar$, which could justify  the saddle point technique but there is no small parameter in the CPI so we cannot use this method. The tools we have in the CPI are only two: 1) the possibility to perform explicitly some integration and 2) the symmetries of $\tilde{\mathcal L}$
which we discovered in \cite{gozzireuter} and which resemble the BRS and anti-BRS transformations of gauge theories. The integration that we can do explicitly in (\ref{FullCPIGeneratingFunctional}) are those in $\lambda_a$ and $\bar c_a$. If we perform them we obtain:
\begin{equation}
\mathcal Z[0]=\int \mathcal D\varphi\mathcal Dc\,\tilde{\delta}\left[\dot{\varphi}^a-\omega^{ab}\frac{\partial H}{\partial \varphi^b}\right]\,\,\tilde{\delta}\left[\dot{c}^a-\omega^{ak}\frac{\partial^2 H}{\partial \varphi^k\partial\varphi^b}\,c^b\right]\label{NewCPI}.
\end{equation}
From this we see that our path-integral forces the paths in $\varphi^a$ and $c^a$ to be only  those which satisfy the eqs.
\begin{align}
&\dot\varphi^a-\omega^{ab}\frac{\partial H}{\partial \varphi^b}=0\\
&\dot c^a-\omega^{ak}\frac{\partial^2 H}{ \partial\varphi^k\partial\varphi^b}c^b=0.\label{PhiCEquations}
\end{align}
These are the first two eqs. of the set represented in (\ref{PhaseSpaceEqn}). How do we get the last two eqs. for $\bar c_a$ and $\lambda_a$?
\section{Ward Identities}
$\tilde{\mathcal L}$ has some BRS and anti-BRS like invariances \cite{gozzireuter} which have the form:
\begin{align}
\textrm{\underline{BRS}:}
\left\{\begin{array}{c} 
\hspace{-0.5cm}\delta\varphi^a=\epsilon\,c^a\\
\hspace{-0.4cm}\delta \bar c_a=i\,\epsilon\,\lambda_a\\
\delta c^a=\delta \lambda_a=0
 \end{array}
 \right.  \textrm{\underline{anti-BRS}:}
 \left\{\begin{array}{cc} 
\hspace{-0.cm}\bar\delta\varphi^a=-i\bar{\epsilon}\,\omega^{ab}\,\bar c_b\\
\hspace{-0.1cm}\bar\delta c^a=i\,\bar{\epsilon}\,\omega^{ab}\,\lambda_b\\
\hspace{-0.2cm}\bar\delta \,\bar c_a=\bar \delta \lambda_a=0\\ 
 \end{array}
 \right.  
\label{BRS}\end{align}
with $\epsilon,\bar \epsilon$ anti commuting parameters. In a path-integral the symmetries can be used to derive what are called the Ward identities. These are derived in the following manner: let us take an arbitrary quantity $O(\varphi,c,\bar c,\lambda)$ and let us do a symmetry variation of it: $\delta O(\varphi,c,\bar c,\lambda)$. It is easy to prove that   
\[
\langle \delta O\rangle =0
\]
where $\langle \dots \rangle$ means an average performed via $\mathcal Z[0]$. The proof of the relation above can be found  in any book on field theory. Let us choose for $O$ the following quantity
\[
O\equiv{\dot\varphi}^{a}-\omega^{ab}\partial_bH
\]
and as symmetry variation the anti-BRS of eq. (\ref{BRS}). We get as Ward identity:
\begin{align*}
\langle \delta O\rangle &\equiv=\langle\delta{\dot\varphi}^{a}-\omega^{ab}\partial_b\partial_dH\bar\delta\varphi^d\rangle=\\
&=-\bar\epsilon\,\omega^{ab}\,\langle\,[\dot{\bar c}_b+\partial_b\partial_dH\,\omega^{ed}\,\bar c_e]\rangle=0,
\end{align*}
which implies 
\begin{equation}
\langle\,[\dot{\bar c}_b+\partial_b\partial_dH\,\omega^{ed}\,\bar c_e]\rangle=0\label{CBarEq}.
\end{equation}
If we now use as a path integral over which to perform the average in (\ref{CBarEq}) the expression (\ref{NewCPI}), which does not depend on $\bar c,$ this variable can be pulled out of the path-integral, while the $\varphi$ entering (\ref{CBarEq}) will sit on the classical trajectory as required by (\ref{NewCPI}). Doing this we get from (\ref{CBarEq}) exactly the equation of motion of $\bar c$:  
\[
\dot{\bar c}_b+\partial_b\partial_dH\,\omega^{ed}\,\bar c_e=0.
\]
This is the manner to get the eq. of motion for $\bar c$ via our path-integral. \par The last equation in  (\ref{PhaseSpaceEqn}) that we have to derive is the one for $\lambda$ . We will  get  also this one  via Ward-identities. Let us choose as $O$ the quantity:
\[
O\equiv\dot c^a-\omega^{ab}\,\partial_b\partial_kH\,c^k
\]
and let us build the $\langle \delta O\rangle$ where the variation we use is the anti-BRS  of (\ref{BRS})
\begin{align}
\langle \delta O\rangle &=\left\langle\left[\bar \delta\dot c^a-\omega^{ab}\partial_b\partial_d(\delta H)\,c^d-\omega^{ab}(\partial_b\partial_dH)\bar\delta c^d\right]\right\rangle\nonumber\\
&=-i\,\overline{\epsilon}\,\omega^{ab}\left\langle \dot{\lambda}_b+i\,\partial_b\partial_d\partial_eH\,\omega^{fe}\bar c_f\,c^d+\right.\nonumber\\
&\left.+\partial_b\partial_dH\,\omega^{ed}\,\lambda_e\right\rangle=0\label{LambdaVariation}.
\end{align}
As before let us choose as generating functional  the one in (\ref{NewCPI}).
As this one does not depend on $\lambda,\,\bar c$, these variables in (\ref{LambdaVariation}) can be pulled out of the average $\langle\dots\rangle$, while the $\varphi,\,c$ will be forced on their classical trajectory by $(\ref{NewCPI})$. The result is:
\[
\dot \lambda_b+i\,\partial_b\partial_d\partial_eH\,\omega^{fe}\bar c_f\,c^d+\partial_b\partial_dH\,\omega^{ed}\,\lambda_e=0,\label{LambdaEq.}
\]
which is the equation of motion for $\lambda$ appearing  in (\ref{PhaseSpaceEqn}). 
\par
What we have done confirms that we do not need to use any least action principle to get the full set of  the $8n$ eqs. of motion but just our path-integral (CPI) and its symmetries.

\section{Equations of motion of the superfields}
The eqs. of motion (\ref{PhaseSpaceEqn}) which we found are equivalent to the following superfield eqs: 
\begin{equation}
\dot \Phi^a=\omega^{ab}\,\frac{\partial H(\Phi)}{\partial \Phi^b}\label{11-1}
\end{equation}
In fact expanding both sides of (\ref{11-1}) in $\theta,\bar{\theta}$ we get:
\begin{align}
\dot\varphi&-\omega^{ab}\partial_bH+\theta\left[\dot c^a-\omega^{ab}\frac{\partial^2H}{\partial\varphi^b\partial \varphi^k}c^k\right]+\label{11-2}\\
&+\bar\theta\,\omega^{ab}\left[\dot{\bar c}_b+\partial_b\partial_dH\,\omega^{ed}\,\bar c_e\right]\nonumber\\
&+i\,\bar \theta\theta\left[\dot \lambda_b+i\,\partial_b\partial_e\partial_lH\,\omega^{fe}\bar c_f\,c^d+\partial_b\partial_dH\,\omega^{ed}\,\lambda_e\right]=0\nonumber
\end{align}
and for this to be zero each coefficient of $\mathbb I,\theta,\bar\theta\theta$ must be zero. These coefficients are exactly the eqs. of motion appearing in (\ref{PhaseSpaceEqn}). So eq.(\ref{11-1}) is a compact way to rewrite the full set of eq.(\ref{PhaseSpaceEqn}). 
By looking at (\ref{11-1}) the reader may think it was derived via  a least action principle from the generating functional appearing in (\ref{SuperCPI})   but there is no reason to use the least action principle  in that  path-integral. We derive the eqs.(\ref{PhaseSpaceEqn}) as we have done before by using only our path-integral and its symmetries and then we sum up all the eqs.(\ref{PhaseSpaceEqn}) like it is done in (\ref{11-2}) via the $\mathbb I,\theta,\bar\theta,\bar\theta\theta$. This sum is zero because each term is zero and this sum is exactly the superfield eq.(\ref{11-1}). So we conclude that also for the superfield eq.(\ref{11-1}) we do not need any least action principle despite their strong resemblance to the Hamilton eq. for $\varphi^a$.

 \section{Conclusions}
In this paper we have derived the eqs. of motion of all the $8\,n$ fields $(\varphi^a,c^a,\bar c_a,\lambda_a)$ appearing in the CPI. We have done so without using neither the least action principle, which has no reason to be in a path-integral formalism, nor the saddle point technique that has no reason to be because we do not have any small parameter in classical mechanics. These $8n$ equations can be summed up into $2n$ compact equations for the superfield $\Phi^a$. The evolution of these super fields is identical to the Hamilton eqs. for $\varphi^a$ but, differently than these, they do not need any least action principle to be derived from. 
The only tools that we used, and that are allowed in our formalism, are the classical path-integral and its associated symmetries.   
\\
\par{\it  Acknowledgements:} The work of E.G and E.C has been financially supported by grants from MIUR (PRIN 2008), INFN (GE41)  and University of Trieste (FRA-2011). 
The work of A. F. Neto has been supported by grants from $\textrm{CNPq}$-Brazil  no.307824/2009-8 and 454357/2011-7.





\bibliographystyle{model1-num-names}
\bibliography{<your-bib-database>}

\begin{thebibliography}{00}
\bibitem{Koopman}
B. O. Koopman, Proc. Natl. Acad. Sci. U.S.A. 17 (1931) 315; J. von Neumann, Ann. Math. 33 (1932) 587.
\bibitem{gozzireuter}
E. Gozzi, M. Reuter, W.D. Thacker, Phys. Rev. D 40 (1989) 3363.
\bibitem{mardsen}
R. Abraham, J. Mardsen, \emph{\bibinfo{title}{Foundation of mechanics}}, Benjamin, New York, 1978.
  \bibitem{geomdeq}
A.A. Abrikosov (jr.), E. Gozzi, D. Mauro, Ann. of Phys. 317 (2005) 24, quant-ph/0406028.

 \end{thebibliography}



\end{document}